\documentclass[12pt]{article}

\usepackage{graphicx}
\usepackage{a4}

\usepackage{fancybox}
\usepackage{epsfig}
\usepackage{color}

\usepackage{amsfonts}
\usepackage{mathrsfs}
\usepackage{amssymb}
\usepackage{amsmath}

\usepackage{dcolumn}
\usepackage{bm}

\def\pa{\partial}

\def\al{\alpha}
\def\be{\beta}

\def\de{\delta}

\def\th{\theta}

\def\la{\lambda}

\def\Ga{\Gamma}

\def\Si{\Sigma}

\def\Om{\Omega}

\newcommand{\ben}{\begin{equation}}
\newcommand{\een}{\end{equation}}
\newcommand{\bea}{\begin{eqnarray}}
\newcommand{\eea}{\end{eqnarray}}
\newcommand{\ba}{\begin{array}}
\newcommand{\ea}{\end{array}}
\newcommand{\bit}{\begin{itemize}}
\newcommand{\eit}{\end{itemize}}

\newcommand{\sgn}{\mathrm{sgn}}

\newcommand{\vs}[1]{\vspace{#1 mm}}

\newcommand{\dsl}{\pa \kern-0.5em /}

\newcommand{\nn}{\nonumber\\}

\begin{document}

\topmargin 0pt \oddsidemargin 0mm

\begin{flushright}

USTC-ICTS-15-14\\

\end{flushright}

\vspace{2mm}

\begin{center}

{\Large

\bf Tachyon field theory description of (thermo)dynamics in dS
space}

\vs{10}

{\large Huiquan Li$^\ast$\footnote{E-mail: lhq@ynao.ac.cn}  and  J. X. Lu$^\star$\footnote{E-mail: jxlu@ustc.edu.cn}}

\vspace{6mm}

{\em

$^\ast$Yunnan Observatories, CAS, Kunming 650011, China

Key Laboratory for the Structure and Evolution of Celestial Objects\\
CAS, Kunming 650011, China\\

and\\

$^\star$Interdisciplinary Center for Theoretical Study\\
University of Science and Technology of China,  Hefei 230026\\

}

\end{center}

\vs{9}

\begin{abstract}
In this work, we present  a few pieces of evidence in support of a possible connection of the gravitational theory
in dS space to the world-volume theory of
unstable D-branes. We show that the action describing the geodesic motion of a massive particle (or
point-like object) in static dS space turns out to be the same as that of  
the tachyon field theory for an unstable particle. The motion along the radial direction from the origin, a locally Minkowski spacetime, to the horizon, a locally Rindler space times a sphere, represents just the tachyon condensation process, therefore providing a geometric picture of tachyon condensation.  We further study a scalar in global or flat dS  and the  tachyon fluctuations in a homogeneous tachyon background, representing either the full or half S-brane, on unstable D-branes and find that certain dynamics of the dS universe  
corresponds to that of the homogeneous full or half S-brane. The thermal temperature of tachyon radiation is found to agree with that felt by any timelike observer in dS space. In string theory context, this temperature is just the Hagedorn one,  signaling a phase transition to closed strings.  An understanding of this transition in the bulk dS space is also given. 
\end{abstract}



\section{Introduction}
\label{sec:introduction}
Space-like or S-branes exist in  various field theories and also in string theory as time dependent solutions\cite{Gutperle:2002ai,Maloney:2003ck}. They are soliton-like configurations localized on a space-like hypersurface and therefore exist only for a moment in time.  In string theory, the space-like cousins of the usual D-branes, called SD-branes, are the ones with the time-coordinate obeying a Dirichlet boundary condition. They can arise as time-dependent solutions of the worldvolume tachyon field theory of an unstable brane, i.e. the non-BPS D-brane or D-brane/antiD-brane pair, and  describe the creation and subsequent decay of this unstable system\cite{Gutperle:2002ai,Maloney:2003ck,Sen:2002nu}.  The original motivation for studying S-branes was to understand holography in the temporal context, following the spatial analog from the usual D-branes.  In the AdS/CFT correspondence, the time-like worldvolume field theory of D-branes holographically reconstructs a spatial dimension.  It is  expected that the Euclidean worldvolume field theory of SD-branes holographically reconstructs a time dimension and this is a necessary ingredient for the proposed dS/CFT correspondence\cite{Strominger:2001pn}. Finding supergravity solutions of these SD-branes served the purpose of  obtaining the de Sitter geometry for the dS/CFT correspondence \cite{Chen:2002yq,Kruczenski:2002ap,Roy:2002ik}. 

In spite of many efforts, this still remains unclear. Recently the near-horizon geometries of SDp-branes were shown to take the form of $(p + 1) + 1$ dimensional de Sitter spaces up to a conformal transformation, upon compactification on $(8 - p)$ dimensional hyperbolic space\cite{Roy:2014mba, Nayek:2014iza} (denoted as $H_{8 - p}$).  However, this is not quite the analog of Dp branes ($p < 5$) for which the corresponding near-horizon geometries are conformal to $AdS_{p + 2} \times S^{8 - p}$ with a linear dilaton and for this reason, unlike the Dp-brane cases, there does not exist a frame such as the so-called `dual frame' \cite{Duff:1991pe,Kanitscheider:2008kd} for which the underlying geometries can be cast exactly as $dS_{p + 2} \times H_{8 - p}$. For non-comformal Dp branes, a holographic renormalization was developed and a so-called generalized holographic conformal structure, giving rise to precision holography, was proposed in \cite{Kanitscheider:2008kd}. For non-conformal SDp branes, we may not expect the same to be true but for low-energy modes, the transverse $H_{8 - p}$ may be decoupled and in certain frame we can have $dS_{p + 2}$. So there might exist some sort of temporal holography for these low energy modes. Similar features had also been argued in \cite{Maloney:2003ck} from a very different perspective. 

Therefore, one might expect a connection of the low energy dynamics of de Sitter space to  that of the worldvolume tachyon field theory of unstable brane, given underlying SD brane relationship to both of them discussed above. We have also other indirect evidence in support of this, mostly in terms of thermodynamics properties.  The mysterious connection of S-branes (or the dynamics of unstable branes), in terms of thermal properties at Hagedorn temperature, to black holes or de Sitter geometries has long been noticed before (see \cite{Maloney:2003ck} for example). For non-extremal black holes/branes being far away from the extremal limit, the corresponding tree-level entropy, while its understanding is still difficult, can be accounted for by the latent heat of a Hagedorn phase transition, whose underlying dynamics is a tachyon condensation process \cite{Dabholkar:2001if}.   This points to us that some forms of time evolution or dynamics of either black holes/branes or dS geometries have connections, one way or the other, to those of tachyon condensation process. The possible connection in certain dynamical properties to the black hole/brane case has recently been investigated in \cite{Li:2011ypa,Li:2013wwa}.  It was found that a probe particle falling in Rindler space, the general near-horizon geometry of non-extremal black hole/brane, can be described by the tachyon field theory for an unstable D-particle. Here the matter falling across horizon process can be equivalently described by that of a tachyon condensation.    

The purpose of this paper is to address analogous issues in the case of de Sitter geometry and to seek the relations between dS space and unstable D-branes in dynamical sense. We shall show in certain cases that the dS (thermo)dynamics is related to that of the corresponding worldvolume tachyon field theory of unstable brane. In Section \ref{sec:SdS}, we
show that the time evolution of a massive probe particle on static dS space or its Euclidean version, i.e. a sphere, can be
precisely described by that of the tachyon field of an unstable D-particle, respectively. This provides a geometric understanding of tachyon condensation process. In
Section \ref{sec:dSU}, we present and discuss certain connections between the linear scalar 
dynamics in expanding dS universe and tachyon fluctuations in a homogeneous tachyon background, representing either full S-brane or future half S-brane, on unstable D-branes. 
In section \ref{sec:therm}, we move to discuss the thermodynamics in both dS space and the tachyon field theory and find that the (Gibbons-) Hawking temperature in dS space agrees with the thermal temperature of tachyon radiation. In string theory context, the temperature of tachyon thermal radiation becomes the Hagedorn one, signaling a phase transition to closed strings. We provide an understanding of this transition in the bulk dS space. 
We conclude this paper  in section 5.

\section{Unstable (D-)particle as probe on static dS patch}
\label{sec:SdS}

For a static dS$_d$ space, the metric is
\begin{equation}\label{e:stadS}
 ds^2=-(1-\be^2r^2)d\eta^2+(1-\be^2r^2)^{-1}dr^2+r^2d\Om_{(d-2)}^2,
\end{equation}
associated with a positive cosmological constant
$\Lambda=(d-1)(d-2)\be^2/2$. The free constant $\be$ will attain a
new meaning in the following discussion when embedded in the string
theory context. The static patch is restricted within the range
$0 \le  r \leq r_H$,  where $r_H\equiv1/\be$ is the radius of the horizon and the horizons are at $r^2 = r_H^2$. Here we focus on  the southern causal diamond
which is defined as $r\in[0,r_H]$, with the south pole located  at $r = 0$. The Hawking temperature of the dS space is:
\begin{equation}\label{e:Hawkingtemp}
 T_{H}^{(\eta)}=\frac{\be}{2\pi}.
\end{equation}

\subsection{Geometric tachyon on the causal static  patch}

With the redefinition,
\begin{equation}\label{e:}
 T=r_H\tanh^{-1}\left(\frac{r}{r_H}\right),
\end{equation}
$ r \in [0, r_H]$ then corresponds to $T \in [0, \infty]$ and in terms of $T$ the metric becomes
\begin{equation}\label{e:redSdSmet}
 ds^2=\frac{1}{\cosh^2(\be T)}\left[-d\eta^2+dT^2+r_H^2\sinh^2(\be T)
d\Om_{(d-2)}^2\right].
\end{equation}
Thus, near the origin $r=0$ or $T=0$, it is a d-dimensional Minkowski
spacetime while  the near-horizon geometry, i.e $r\rightarrow  r_H$ or
$T\rightarrow \infty$,  is the Rindler spacetime times a (d -2)-sphere. Note that the above metric
 interpolates these two.

We now consider a test particle (or point-like object) with mass
$\tau_0$ moving along the radial direction in the above background (\ref{e:redSdSmet}) . The geodesic motion of this particle can be described by the following action 
now
\begin{equation}\label{e:stadstac}
 S_0=-\int d\eta V(T)\sqrt{1-\dot{T}^2}, \textrm{ }\textrm{ }\textrm{ }
V(T)=\frac{\tau_0}{\cosh(\be T)}
\end{equation}
where the dot denotes the derivative with respect to $\eta$. 

This is exactly the action \cite{Garousi:2000tr,Bergshoeff:2000dq}
for a non-BPS D-particle (or homogeneous non-BPS D$p$-brane) with the tachyon field $T$ and  the correct
tachyon potential $V(T)$ \cite{Kutasov:2003er,Smedback:2003ur} as derived in
open string field theory (Now $T$
attains the new meaning of a scalar field in this action).  In other words, the radial geodesic motion of a massive particle in the causal diamond in static dS patch is nothing but the dynamics of the tachyon field of an unstable D-particle.  This is the first example presented in connecting the gravitational theory in dS space to the worldvolume theory of unstable branes.   In addition, this provides also a  realization of the so-called geometric tachyon \cite{Kutasov:2004dj,Kutasov:2004ct,Li:2011ypa}. In order to bring the above connection, we need to have the constant $\be=1/(2l_s)$ for bosonic strings and
$\beta =1/(\sqrt{2}l_s)$ for superstrings, with the string length
$l_s=\sqrt{\al'}$. Either gives the corresponding correct Hagedorn temperature, respectively.  One may question the validity of the geometry (\ref{e:redSdSmet}) at such a high temperature but  the obtained Hawking temperature nevertheless agrees with the one obtained by more sophisticated stringy computations\cite{Strominger:2002pc,Maloney:2003ck}.  We have to admit  that an understanding of this agreement is still lacking.  

Let us take a close look of the connection between the dynamics in dS space and that of tachyon field of the unstable brane.
From the tachyon potential given in (\ref{e:stadstac}),  one can see that $T = 0$ ($r = 0$) gives the maximum of the potential, therefore corresponding to the open string vacuum, while $T = \infty$ ($r = r_H$) gives the minimum, corresponding to the closed string vacuum. In the former case, the geometry of metric (\ref{e:redSdSmet}) is a Minkowski spacetime and this is consistent with the open string vaccum and in the latter the closed string vaccum is the Rindler spacetime times a sphere discussed above. So the tachyon rolling from the top of the potential to the closed string vacuum can be viewed as a geodesic motion from $r = 0$ (the south pole) to the horizon at $r = r_H$. So we provide geometric pictures for the two vacua and the tachyon rolling process in the tachyon field theory of unstable D-brane, respectively.

The underlying geometric picture may provide, borrowed from the gravity side, certain understanding of the tachyon condensation, especially towards the end of this process.  From the tachyon action (\ref {e:stadstac}), we have the energy-momentum tensor  
\begin{equation}\label{e:statensor}
 T_{00}=\frac{V(T)}{\sqrt{1-\dot{T}^2}}=E,
\textrm{ }\textrm{ }\textrm{ } T_{ij}=-\frac{V^2}{E}\de_{ij}.
\end{equation}
The energy is conserved mainly due to the classical limit $g_s \rightarrow 0$ taken in the effective description (So the closed string radiations are ignored)\footnote{We will confine ourselves in this paper to this classical discussion, not touching the subtle $g_s \rightarrow 0$ as well as other issues, for example, raised in \cite{Larsen:2002wc}. However, when we consider thermodynamics, the appearance of Hagedorn temperature nevertheless signals a transition to closed strings.}. From the first equation in (\ref{e:statensor}), mathematically the tachyon field will accelerate to the critical value $|\dot{T}|=1$ in the limit of $T \rightarrow \infty$, i.e., reaching the speed  of light and this appears also consistent with the geometric picture of geodesic at the horizon. The pressure in this limit tends to zero, giving the so-called pressureless tachyon matter \cite{Sen:2002in,Sen:2002an,Sugimoto:2002fp,Buchel:2002tj}\footnote{We thank the the anonymous referee for bringing the useful references \cite{Sugimoto:2002fp,Buchel:2002tj} to our attention.}.  However, the classical limit $g_s \rightarrow 0$ and the end products among other things were questioned for example in \cite{Larsen:2002wc}.  From the geometric side,  when a geodesic motion is considered, to an observer or detector at the south pole at $r = 0$, the horizon location at $r = r_H$ and the coordinate time $\eta = \infty$ are not well-defined and are usually defined only as a limiting process.  Recent study indicated that at least for black holes, when a probe passes through the horizon, the location of the horizon is actually not fixed but changes \cite{Parikh:1999mf} and this will be the key for accounting for the information conservation as demonstrated in a series of publications \cite{Zhang:2009jn,Zhang:2009td,Zhang:2013sza,Zhang:2013win,Guo:2014bza,Hui:2014hpa}.  Related considerations in spirit have also been given very recently in \cite{Hawking:2015qqa, Strominger:2014pwa}. In the present context, we expect that similar things may also occur and so the limit $r \rightarrow r_H$ and $\eta \rightarrow \infty$ may not be taken smoothly as usually done. Here new physics will enter in the process. Transforming this to the process of tachyon condensation implies that from the open string vacuum at the top of the potential to the closed string one at the bottom of the potential may not be a smooth one and quantum effects should be considered, where the appearance of Hagedorn temperature is an indication of this when thermodynamics is considered.   

\subsection{Geometric tachyon on a sphere}

Now we consider a probe moving along the radial direction on the Euclidean version of dS geometry\footnote{Note that the metric for the $(d - 2)$-sphere in (\ref{e:redSdSmet}) is now irrelevant for this motion.} .  This amounts to taking the time in (\ref{e:redSdSmet}) to be an Euclidean one, i.e.,  by sending $\eta\rightarrow-i\tau$ there and then $\beta \tau \rightarrow \varphi$ ($\varphi\in[0, 2\pi]$) and $\beta T \to T$ with $\beta = 1/r_H$.  We have then the relevant geometry 
\begin{equation}\label{e:}
 ds_S^2 = \frac{r_H^2}{\cosh^2T}(d\varphi^2+dT^2).
\end{equation}
If we further extend $T$ from $0 \le T < \infty$ to  $- \infty < T < \infty$, the above geometry represents then a 2-sphere and can be cast as 
\begin{equation}\label{e:2-sphere}
 ds_S^2 = r_H^2 \left( \cos^2\al d\varphi^2+d\al^2\right) = r_H^2\left(\sin^2\th d\varphi^2+d\th^2 \right), 
\end{equation} 
where we have introduced angles $\theta$ and $\alpha$ related by  $\al=\pi/2-\th$ with $\cos\al=1/\cosh T$.  Here $T=0$ gives $\al=0$, $T = \infty$ corresponds to $\al = \pi/2$ and  $T = - \infty$ to $\al = - \pi/2$. So we have now $\alpha\in[- \pi/2, \pi/2]$ and $\th\in[0,\pi]$.  

The geodesic motion for the probe on the 2-sphere can  equivalently be described by the following Euclidean 
tachyon field theory
\begin{equation}\label{e:eaction}
 S^E_0=\tau_0 L= r_H \int d\varphi V(T)\sqrt{1+{T'}^2},
\textrm{ }\textrm{ }\textrm{ } V(T)=\frac{\tau_0}{\cosh T},
\end{equation}
where $T'=\pa_\varphi T$ and $L$ describes distance on the sphere.
The field equation can be integrated once to give
\begin{equation}\label{e:ee}
 \frac{V}{\sqrt{1+{T'}^2}}=V_0.
\end{equation}
Here the constant $V_0=V(a)$ with $a$ determined by $T'|_{T=a}=0$. The solution to the above equation determines the shortest distance on the sphere.

When $V_0=\tau_0$, the system is on the top of the potential. From
the above equation (\ref{e:ee}), we have the only possible solution  $T=0$ and
$T'=0$. In other words, the tachyon is fixed on
the equator $\th=\pi/2$. In general $0 <  V_0 \le \tau_0$ and keeping in mind the aforementioned extension, we can have the explicit solution from (\ref{e:ee}) as \begin{equation}\label{e:kinksol}
 T(\varphi)=\sinh^{-1}\left(\sqrt{\frac{\tau_0^2}{V_0^2}-1}
\sin\varphi\right).
\end{equation}
This is the kink-anti-kink tachyon solution
\cite{Kim:2003in,Brax:2003rs,Hindmarsh:2009hs}, with kink and
anti-kink located at $\varphi= 0$ and $\varphi= \pi$ (which can be extended to $\varphi = 2 N \pi$ and $\varphi = (2 N + 1)\pi$ with $N\in Z$), respectively. Since these positions are periodically identified
with a period $2\pi$, there are actually $N$ coincident D-banes at
the kink position and $N$ coincident anti-D-branes at the anti-kink
position.

Using the spherical coordinates $(\theta, \varphi)$, this solution is 
\begin{equation}\label{e:grecir}
 \sin\th=\frac{V_0}{\sqrt{V_0^2\cos^2\varphi
+\tau_0^2\sin^2\varphi}}.
\end{equation}
This describes a family of great circles labelled by the parameter $V_0$, with each passing through the two points at $\varphi = 0$ and $\varphi = \pi$ on the 2-sphere,  starting from
the equator at $\th=\pi/2$ and approaching the vertical one when the $V_0$ decreases from $\tau_0$ to $0$. 

The static energy of the system is the mass times the
geodesic  distance on the sphere  and is given as
\begin{equation}\label{e:}
 E= \frac{1}{\beta} \int d\varphi V\sqrt{1+{T'}^2}=\frac{1}{V_0}\int
d\varphi V^2=\tau_0 L.
\end{equation}
Note that the geodesic distance of each great circle on the 2-sphere of unit radius is always $2 \pi$, which can be calculated directly from the above by using the solution given in (\ref{e:grecir}).  So we have $L =  2\pi/\beta$, independent of $V_0$, and this is consistent
with the analysis in \cite{Brax:2003rs}. So the energy for every
half period (because there are a kink and an anti-kink within a
period) is
\begin{equation}\label{e:}
 E_{\frac{1}{2}}=\pi\tau_0/\beta.
\end{equation}
If we replace $\tau_0$ with $\tau_p$ for a homogeneous non-BPS D$p$-brane and construct
the potential $V(T)=\tau_p/\cosh(\be T)$, the energy becomes
$E_{1/2}=\pi\tau_p/\be$, which gives rise to the tension of a
(anti-)D$(p-1)$-brane if $\beta$ is given by the corresponding string length $l_s$ as discussed in the previous subsection. 

\section{dS universe and S-brane}
\label{sec:dSU}

If we want to consider anything more than that covered by the southern (or northern) causal diamond, we enter into a dS universe in other coordinates. 
In this section, we shall show that the dynamics in
$d$-dimensional dS universe\footnote{We limit ourselves to 
the non-trivial cases of $d \ge 2$ from now on.} in the global or flat coordinates is relevant to a decaying unstable
D$p$-brane with $d=p+1$.

We consider first the possible homogenous tachyon backgrounds relevant and the bosonic part of the DBI effective action of a non-BPS D$p$-brane
in flat Minkowski metric as
\cite{Garousi:2000tr,Bergshoeff:2000dq,Sen:2004nf}
\begin{equation}\label{e:fulltacact}
 S_p=-\int d^{p+1}xV(T)\sqrt{-\det(\eta_{\mu\nu}+\pa_\mu T\pa_\nu T
+ \cdots)},
\end{equation}
where $\cdots$ stand for the omitting transverse scalars and gauge fields which are irrelevant for the present discussion. The tachyon potential can
be taken as
\begin{equation}\label{e:Dppot}
 V(T)=\frac{\tau_p}{\cosh(\be T)} \,,
\end{equation}
where $\tau_p$ is the tension of the non-BPS D$p$-brane. 

For the homogenous case, the tachyon background $T_0 (t)$ can be found from the above action as follows. The 
non-vanishing components of the energy-momentum tensor from the
action (\ref{e:fulltacact}) are
\begin{equation}\label{e:homenmotensor}
 T_{00}(T_0)=\frac{V(T_0)}{\sqrt{1-\dot{T}_0^2}}=E,
\textrm{ }\textrm{ }\textrm{ }
T_{ij}(T_0)=-\frac{V^2(T_0)}{E}\de_{ij},
\end{equation}
where $E$ is the conserved energy of the system in the limit
$g_s = 0$. Defining $l\equiv E/\tau_p$, we can have three different types of time-dependent solutions, depending on $l > 1,  l = 1$ and $l < 1$.  For each of the three cases, we can have more general solutions with different $ T_0 (0)$ and $\dot T (0)$ at $t = 0$ but with the same $T_0 (\pm \infty)$ and $\dot T_0 (\pm \infty)$.  In other words, these general solutions have the same characteristic behavior as the simplest solutions discussed below. For simplicity, we will focus only on these simplest ones in what follows. \\ 

\noindent
{\bf Case  $l > 1$:} the simplest solution with $\dot T_0 (t) |_{t \to \pm \infty} = 1$ is\footnote{We can also have solutions of $\dot T_0 (t\to \pm \infty) = -1$ and this describes $T_0 (- \infty) = \infty$ to $T_0 (
+ \infty) = - \infty$ as $t = - \infty$ to $t  = + \infty$, just the reverse of the $\dot T_0 (t \to \pm\infty)= 1$ process discussed in the text.  So we will not repeat this case here.}  
\begin{equation}\label{e:T0sol}
 T_0(t)=\frac{1}{\beta}\sinh^{-1}\left[\sqrt{1-\frac{1}{l^2}}
\sinh{(\beta t)}\right],
\end{equation}
\begin{equation}\label{e:dotT0sol}
 \dot{T}_0 (t) =\sqrt{\frac{l^2-1}{l^2-\tanh^2{(\beta t)}}}.
\end{equation}
Thus, $T_0 (0) =0$, $\dot{T}_0 (0) =\sqrt{1-1/l^2}$ at $t=0$ and
$T_0 (t) \rightarrow\pm\infty$, $\dot{T}_0 (t) \rightarrow1$ as
$t\rightarrow\pm\infty$, respectively.  This describes brane creation and then decay process, i.e., a full S-brane process \cite{Strominger:2002pc}.\\

\noindent
{\bf Case $l = 1$:} we can have the following solutions
\begin{equation}\label{lone}
\begin{array}{ccc}
T_0 (t) = \frac{1}{\beta} \sinh^{-1} \left[\lambda \, e^{\beta t}\right] &    \qquad\qquad\qquad     &  T_0 (t) =  \frac{1}{\beta} \sinh^{-1} \left[\lambda\, e^{- \beta t}\right] \\
       &   {\rm or} &    \\
\dot T_0 (t) = \frac{\lambda \, e^{\beta t}}{\sqrt{1 + \lambda^2 e^{2 \beta t}}} \qquad\qquad&\qquad \qquad\qquad  & \dot T_0 (t) = \frac{- \lambda \, e^{- \beta t}}{\sqrt{1 + \lambda^2 e^{- 2 \beta t}}}, \qquad\qquad \end{array}
\end{equation}
where we have chosen the parameter $\lambda > 0$ ($\lambda < 0$ gives the other set of solutions with similar properties).  For  $t \to  - \infty$, the above first set of solutions give 
$T_0 (- \infty) \to 0,  \dot T (- \infty) \to  0$ but the second set give $T_0 (- \infty) \to + \infty,  \dot T_0  (- \infty) \to - 1$. For $t \to \infty$,  the first set give $T_0 (\infty) \to \infty, \dot T_0 (\infty) \to 1$ while the second set give $T_0 (\infty) \to 0, \dot T_0 (\infty) \to 0$.  So this says that the first set of solutions describe the brane decay while the second set describe
the brane creation, which describes future and past  half of an S-brane, respectively, according to \cite{Strominger:2002pc,Maloney:2003ck}. \\

\noindent
{\bf Case $0 < l < 1$:} we have now the simplest solutions,
\bea \label{smalllbg}
T_0 (t) &=& \frac{1}{\beta} \sinh^{-1} \left[\sqrt{\frac{1}{l^2}  - 1} \, \cosh \beta t\right], \nn
\dot T_0 (t) & = & \frac{\sqrt{1 - l^2} \tanh \beta t}{\sqrt{1 - l^2 \tanh^2 \beta t}}.
\eea
 From the above, one can see that as $t \to \pm\infty$,  $T_0 (\pm \infty) \to + \infty$ while $\dot T_0 (\pm \infty) \to \pm 1$, respectively.  This represents brane creation and then brane decay into the same vacuum, therefore also a full S-brane. This case differs from the $l > 1$ one for which the brane creation from one vacuum and then decay into the other one. Note here that $\dot T (0) = 0$ but with a finite $T (0) > 0$ at $t = 0$, a consistent picture for $l < 1$.

\subsection{The global dS and the full S-brane}
The dS universe in global coordinates is 
\begin{equation}\label{e:closeddSu}
 ds^2=-dt^2+r_H^2\cosh^2(\be t)d\Om_{(d-1)}^2,
\end{equation}
with $r_H=1/\be$. The metric is invariant under the time reflection
$t\rightarrow-t$ . 

We shall compare dynamics in dS universe
(\ref{e:closeddSu}) with tachyon fluctuations on a homogeneously evolving
tachyon $T=T_0(t)$ background which is a solution of the first equation in (\ref{e:homenmotensor}).

For this, let us first consider the Klein-Gordon (KG) equation for a scalar with mass $m_s$ on the
dS geometry as
\begin{equation}\label{e:KGclodS}
 \left[\pa_t^2+(d-1)\be\tanh(\be t)\pa_t-\frac{1}
{r_H^2\cosh^2(\be t)}\nabla_S^2+m_s^2\right]
\phi(t,\Om)=0.
\end{equation}

The above equation can be simplified, by setting $\phi=\cosh^{-(d-1)/2}(\be t)\hat{\phi}$, to 
\begin{equation}\label{e:standKGclodS}
 \left[\pa_t^2+m_s^2-\frac{(d-1)^2}{4r_H^2}+\frac{1}
{r_H^2\cosh^2(\be t)}\left(\frac{(d-1)(d-3)}{4}
-\nabla_S^2\right)\right]\hat{\phi}=0.
\end{equation}

We now turn our attention to the effective action (\ref{e:fulltacact}) for pure
tachyon field as
\begin{equation}\label{e:puretacact}
 S_p=-\int d^{p+1}xV(T)\sqrt{1+\eta^{\mu\nu}\pa_\mu T\pa_\nu T}.
\end{equation}
The equation of motion for the tachyon from the above action is 
\begin{equation}\label{e:Teom}
 \left(\pa_\mu\pa^\mu T-\frac{V'}{V}\right)(1+\pa T\cdot\pa
T)=\frac{1}{2}\pa^\mu T\pa_\mu(1+\pa T\cdot\pa T),
\end{equation}
where the prime stands for the derivative with respect to $T$. We
now consider  tachyon perturbations  to a homogeneous
tachyon background: $T(t,\vec{x})=T_0(t)+\tau(t,\vec{x})$. To
leading order in the perturbation, we have the equation as
\begin{equation}\label{e:per}
 \left[\ddot{T}_0-\be\tanh(\be T_0)(1-\dot{T}_0^2)\right]+
\left[\ddot{\tau}+ 2\be\dot{T}_0\tanh(\be
T_0)\dot{\tau}-(1-\dot{T}_0^2)
\vec{\nabla}^2\tau-\frac{\be^2(1-\dot{T}_0^2)}{\cosh^2(\be
T_0)}\tau\right]=0.
\end{equation}
If $T_0 (t)$ is a tachyon background satisfying the first equation in (\ref{e:homenmotensor}), as given previously,
the first square bracket in the above equation vanishes and we have then
\ben\label{pe}
\ddot{\tau}+ 2\be\dot{T}_0\tanh(\be
T_0)\dot{\tau}-(1-\dot{T}_0^2)
\vec{\nabla}^2\tau-\frac{\be^2(1-\dot{T}_0^2)}{\cosh^2(\be
T_0)}\tau = 0.
\een
We can also simplify the above equation by setting $\tau (t) = \hat\tau (t)/\cosh\beta T_0 (t)$ and end up with the following equation 
\ben\label{spe}
\left[\partial_t^2 - \beta^2 - \frac{1}{l^2 \cosh^2\left(\beta T_0\right)} \vec{\nabla}^2\right] \hat\tau = 0.
\een
In obtaining the above, we have made use of the first equation in (\ref{e:homenmotensor}) for the tachyon background $T_0 (t)$. 

We now try to examine, under what conditions, that the equation (\ref{spe}) can be identified with (\ref{e:standKGclodS}). To be so, first the non-BPS Dp-brane has to  wrap on a p-sphere with a constant radius $r_0$ such that $\vec{\nabla}^2 = \nabla_S^2 /r_0^2$.  Secondly, we need to have $\cosh^2 \beta T_0 (t) \sim \cosh^2 \beta t$. This later condition can only be met for  $l \gg 1$ or $l \ll 1$ case discussed earlier.  Let us examine each to see which is the proper one needed. For $l > 1$ case, we have from  (\ref{e:T0sol}) 
\ben \cosh^2 \beta T_0 (t)  = \cosh^2 \beta t - \frac{1}{l^2} \sinh^2 \beta t  \approx \cosh^2 \beta t,\een where the last approximation holds only if $l \gg 1$. However, there exist various issues for this case. First for $l \gg 1$ or $E \gg \tau_p$,   the system energy is much larger than the tachyon potential height and one expects that the tachyon background $T_0 (t)$ changes dramatically in the course of time evolution from $t \to - \infty$ to $t \to + \infty$.  So it is hardly possible to keep the fluctuations of this background small in the process and therefore this will not be suitable to the expected case discussed in the Introduction for which the relevant excited modes should be low energy ones.   Secondly, in order to identify (\ref{spe}) with  (\ref{e:standKGclodS}), we need further to have $l r_0   = r_H$. This says that $r_0 = r_H/l \ll r_H$ when $l \gg 1$. So the p-brane wrapped p-sphere has a radius much smaller than the p-sphere radius in dS geometry when $l \gg 1$, therefore giving a very large curvature. This will cause the DBI effective action
(\ref{e:fulltacact}) invalid in our above discussion\footnote{In our above discussion, we change the original flat Minkowski spacetime to $R \times S^p$ with the p-sphere a constant radius $r_0$ for the worldvolume.  This should be straightforward and no complications should arise.} .  Let us now move to the $l < 1$ or $E < \tau_p$ case for which we have from 
 (\ref{smalllbg})
 \ben \cosh^2 \beta T_0 (t) = \frac{1}{l^2} \cosh^2\beta t - \sinh^2 \beta t \approx   \frac{1}{l^2} \cosh^2\beta t,\een
 where the last approximation holds if $l \ll 1$.  So to have $\cosh \beta T_0 (t) \propto \cosh\beta t$, we need $l \ll 1$ or $E \ll \tau_p$.  In other words, the system energy is much smaller than the tachyon potential height and for this, one doesn't expect the background to change dramatically in the course of time evolution from $t \to - \infty$ to $t \to + \infty$.  
When $l \ll 1$, from the above and (\ref{smalllbg}), we have $l\sinh \beta T_0 (t) \approx  l \cosh \beta T_0 (t) \approx \cosh \beta t > 1$. So for any time, we should have a large $T_0 (t) \gg 1$.  In other words, the system is always nearby its vacuum at $T (t) = \infty$ for $t\to -\infty$ to $t \to + \infty$. This lends further support to the above slow-change background claim. Moreover, this is also good for the worldvolume geometry since for present we need only $r_0 = r_H$, not encountering the issue as discussed above for $l \gg 1$.  One also expects that the relevant modes are low energy for this case.  So $l \ll 1$ is the focus of our further discussion.  

For this,  the equation (\ref{spe}) becomes now
\ben \label{ftaue}\left[\partial_t^2 - \beta^2 - \frac{1}{r^2_H \cosh^2\left(\beta t\right)} \nabla_S^2\right] \hat\tau = 0,
\een
where we have set $r_0 = r_H$ (Note also $\beta = 1/r_H$).  So we are now ready to make possible identification of the above equation with  (\ref{e:standKGclodS}).  For $d = 3$, the two equations can be identified if $m_s^2 = 0$.  For this case, the probe scalar is massless.  In general ($d \neq 3$), it doesn't appear that we can identify the two equations. Fortunately, the spatial spaces for both cases are p-sphere, so we have $\hat \phi (t, \Omega) = f_{m_s, r_H, p, l_1} (t) Y^{(l_1)} (\Omega)$ and $\hat \tau (t, \Omega) = g_{r_H, r_0, p, l_2} (t) Y^{(l_2)} (\Omega)$ with $\nabla_{S^p}^2 Y^{(l_i)}  (\Omega) = - l_i(l_i + p - 1) Y^{(l_i)} (\Omega)$ with $l_i = 0, 1, \cdots$ for $i = 1, 2$, respectively. Here $Y^{(l)} (\Omega)$ are the p-sphere scalar spherical harmonics. Note  that we still leave the radius of p-brane wrapped sphere as $r_0$ for the purpose of more options. With these, equation (\ref{e:standKGclodS}) becomes 
 \ben\label{ffe}
 \left[\pa_t^2+m_s^2-\frac{(d-1)^2}{4r_H^2} + \frac{1}
{r_H^2\cosh^2(\be t)}\left( l_1(l_1 + d - 2) + \frac{(d-1)(d-3)}{4}
\right)\right] f_{m_s, r_H, p, l_1}  (t) =0,
 \een
 while equation (\ref{ftaue}) becomes 
  \ben\label{fge}
  \left[\partial_t^2 - \beta^2 + \frac{l_2(l_2 + d - 2)}{r^2_0 \cosh^2(\beta t)} \right] g_{r_H, r_0, p, l_2 } (t) = 0.
  \een
  In the above, we have set $p = d - 1$. Then the above two equations can be identified if the following equations hold.
    \ben 
  m_s^2 = \frac{(d-3)(d + 1)}{4r_H^2}, \qquad\qquad   l_1 = \left[\left(\frac{r_H}{r_0}\right)^2 l_2 (l_2+ d - 2)  + \frac{1}{4}\right]^{1/2} - \frac{d - 2}{2},\een
  where $\beta = 1/r_H$ has been used.
  For $d = 2$,  we have $m_s^2 < 0$ and for $d > 3$, $m_s^2 > 0$.  There is no issue for the first equation to hold always.  However, it is quite non-trivial for the second one to hold since both $l_1$ and $l_2$ are non-negative integers.   Note in general $2 \le d \le 10$.  As discussed above, for $d = 3$, we can have $l_1 = l_2$ for $r_0 = r_H$, therefore there is no restriction from the second equation above. For other $d$, having this equation to hold is highly non-trivial. If we take $r_0 = r_H$, there exist possible solutions only for odd $d$, i.e., $d = 5, 7, 9$ cases.   For proper choice of $r_0$, we do have solutions for all $d$ allowed but their existence are rare in general. Let us take $d = 2, 4, 5$ as three examples. We will consider the range  of $l_2$ as $1\le l_2 \le 10^4$. For $d = 2$ case, $m_s^2 =   - 3/4 r_H^2$, a tachyon in the sense of flat spacetime. We now have 
 \ben \label{d2case}
 l_1   = \left[\left(\frac{r_H}{r_0}\right)^2 l_2^2 + \frac{1}{4}\right]^{1/2}.
 \een 
 In the above, $l_2 = 0$ can never be a solution, neither does $l_1 = 0$. If we want $l_2 = 1 = l_1$ to be solution, then we need to take $r_0 = 2 r_H /\sqrt{3}$.  There exist only four pairs\footnote{We thank Xiaojun Tan for help on finding the pairs of solutions for the three cases considered.} of solutions $(l_1, l_2)$, which are $(1, 1), (13, 15), (181, 209)$ and $(2521, 2911)$ for the range of $l_2$ mentioned above.  For $d = 4$ case, we have $m_s^2 = 5/(4 r_H^2) > 0$ and 
 \ben\label{d4}
 l_1 = \frac{1}{2} \left[5 l_2 (l_2 +  2) + 1\right]^{1/2} - 1,
 \een
 where we have taken $r_H/r_0 = \sqrt{5}/2$. There exist now only three pairs of solutions $(l_1, l_2)$: $(1, 1), (37, 33)$ and $(681, 609)$ for the allowed range of $l_2$. For the $d = 5$ case, we have  $m_s^2 = 3/r_H^2  > 0$ and 
 \ben\label{d5case}
 l_1 = \left[\frac{3}{2} l_2 (l_2 + 3) + \frac{1}{4}\right]^{1/2} - \frac{3}{2},
 \een
 where we have chosen $r_H/r_0 = \sqrt{3/2}$. There exist now twelve pairs of solutions $(l_1, l_2)$ for the allowed range of $l_2$, which are $(1, 1), (5, 4), (10, 8), (26, 21), (64, 52), (113, 92), (271, 221)$, $(647, 528), (1132, 924), (2696, 2201), (6418, 5240)$ and $(11219, 9160)$.  For different $d \neq 3$ and for different choice of $r_H/r_0$, the number of pairs of solutions $(l_1, l_2)$ are different but in general not so many.  For now, we don't understand why the solutions are so rare.  One possible reason for this may be that the probe scalar satisfying its equation  (\ref{e:standKGclodS}) can be either future or past causally connected to an observer, for example, in the south pole while the tachyon perturbation needs to be in the southern causal diamond.  The requirement of the two having the same solutions may put very strong restriction on the solutions or this provides a selection rule.

\subsection{The flat dS universe and the half S-brane}
\label{sec:FdSU}
The flat dS universe metric is
\begin{equation}\label{e:flatdSu}
 ds^2=-dt^2+e^{2\be t}d\vec{x}^2_{(d-1)},
\end{equation}
where the spatial part of metric is a flat $(d - 1)$-dimensional Euclidean space. It
describes an expanding universe with time running from $0$ to
$\infty$. The contracting dS universe with the exponential factor
$e^{-2\be t}$ can be viewed as its inverse process.

The K-G equation of a probe scalar $\phi$ with mass $m_s$ in this spacetime background is
\begin{equation}\label{e:KGeom}
 \left[\pa_t^2+(d-1)\be\pa_t+m_s^2-e^{-2\be t}\vec{\nabla}^2
\right]\phi(t,\vec{x})=0.
\end{equation}
The above can be simplified if we set $\phi=e^{-(d-1)\be t/2}\hat{\phi}(t,\vec{x})$. We then end up with
\begin{equation}\label{e:redKGeom}
 \left[\pa_t^2-\frac{1}{4}(d-1)^2\be^2+m_s^2
-e^{-2\be t}\vec{\nabla}^2\right]\hat{\phi}(t,\vec{x})=0.
\end{equation}

The relevant corresponding tachyon background should be the first solution of (\ref{lone}), a future half S-brane, for $l = 1$ case.  In other words, we have $\sinh \beta T_0 (\sigma^0) =  \lambda e^{\beta \sigma^0}$, where for the following purpose, we denote the worldvolume coordinates as $\sigma^\mu$ with $\mu = 0, 1, \cdots p$ and with $\sigma^0$ the worldvolume time coordinate. We also consider here time $\sigma^0$ running from $0$ to $\infty$ as the above expanding dS universe.   To have the tachyon background 
$T_0 (\sigma^0)$ to be large as before, i.e., nearby the vacuum at $T_0 = \infty$,  we need to have a large $\lambda$ for the present case. So we have, for large $\lambda$,
\ben 
\cosh^2 \beta T_0 (\sigma^0) = 
1 + \lambda^2 e^{2 \beta \sigma^0} \approx \lambda^2 e^{2\beta \sigma^0}.\een
    We then have from (\ref{spe}) for the tachyon fluctuation for the present case as
\ben\label{frtf}
\left[\partial_t^2 - \beta^2 - \frac{1}{\lambda^2} e^{- 2 \beta t} \vec{\nabla}_{\vec \sigma}^2\right] \hat\tau (t, \vec\sigma) = 0,
\een
where we have set $\sigma^0 = t$. Then both (\ref{e:redKGeom}) and (\ref{frtf}) can be identified if we set $x^i = \lambda \sigma^i$ with $i =1, \cdots, p$ and the scalar mass is given by
$m_s^2=(d-3)(d+1)\be^2/4$ for any allowed $d$.

\section{Tachyon description of thermodynamics in dS space}
\label{sec:therm}
In the previous sections, we have demonstrated that the low energy dynamics of dS space is closely related to that of tachyon field theory.  The detail of this connection depends on 
the coordinate system used to describe the dS space, which gives different cover of dS space as well as different insights into the dS structures.  This may in some sense reflect the fact that a single immortal observer in de Sitter space can see at most half of the space and therefore a description of the entire space goes beyond what can be physically measured.

As discussed in the Introduction, it has long been noticed that the open string vacua on S-brane have somewhat mysterious thermal properties reminiscent of dS vacua \cite{Strominger:2002pc,Gutperle:2003xf}. This has been demonstrated using the minisuperspace approximation in \cite{Maloney:2003ck}.  Given the connection of S-brane to the tachyon condensation of unstable D-brane and the tachyon itself as the open string vacuum, one expects that the tachyon effective theory itself can spell out the thermal properties such as the Hagedorn temperature. This temperature, as will be demonstrated, is actually the Hawking or Gibbon-Hawking temperature related to the (cosmological) event horizon of dS space,  therefore this provides also a support to the thermal connection between dS space and the tachyon effective theory.  

\subsection{Thermal radiation in tachyon effective field theory}

Open-string pair creation of thermal spectrum has been revealed in
the tachyon BCFT
\cite{Strominger:2002pc,Gutperle:2003xf,Maloney:2003ck}, in which
the free worldsheet action is deformed by the boundary term
\cite{Sen:2002nu,Sen:2002in}:
\begin{equation}\label{e:BCFTaction}
 S_{\textrm{bndy}}=\int_{\pa\Si}d\tau m^2(X^0(\tau)).
\end{equation}
The theory with $m^2=(\la/2)e^{\pm 2\be X^0}$ respectively describes
the decay/creation process of an unstable D-brane, i.e., half
S-brane. The one with the combination $m^2=(\la/2)e^{\pm 2\be X^0}$
describes the full S-brane.

By considering  only the zero modes $(x^0,\vec{x})$ of
$(X^0,\vec{X})$ in the theory, the authors in \cite{Maloney:2003ck} derived that, during the
tachyon condensation process, open-string pairs with a thermal
spectrum can be produced at the Hagedorn temperature
\begin{equation}\label{e:Hagedorntemp}
 T_{\textrm{Hag}}^{(x^0)}=\frac{\be}{2\pi}.
\end{equation}
This temperature implies that the rolling tachyon will decay into
closed strings.

The tachyon effective action (\ref{e:stadstac}) or (\ref{e:puretacact}) for the homogeneously
decaying case is believed to be equivalent to the time-evolving
tachyon BCFT \cite{Okuyama:2003wm,Lambert:2003zr,Sen:2004nf}.
In \cite{Li:2013wwa},
it has been shown that the thermal feature of particle creation with the
same temperature for black holes  (\ref{e:Hagedorntemp}) can also be derived  in
the effective theory.

In this subsection, we will use the tachyon effective theory to study the particle creation during the course of tachyon condensation and to obtain the temperature of the corresponding  thermal radiation for the purpose of comparison with the thermal temperature of dS space first without reference to string theory.  This will lend further support of the proposed connection between  the two thermodynamically.  We then discuss the implications of this in the context of string theory, especially on the Hagedorn transition. 

   We intend to have the Klein-Gordon equation for the tachyon wave function. The tachyon DBI
effective action itself (\ref{e:stadstac}) or (\ref{e:puretacact}) cannot serve this purpose because of its non-linearity. So we seek an alternative for this purpose. We are looking for particle 
creation during the course of tachyon condensation. In other words, we are considering the particle creation for tachyon satisfying the first equation of  (\ref{e:statensor}) or (\ref{e:homenmotensor}), i.e., 
\begin{equation}\label{e:effEoM}
 \dot{T}^2= 1 - \frac{V^2}{E^2},
\end{equation}
where the `dot' over $T$ represents the time derivative of $T$ (the time is the worldvolume one which can be identified with the respective dS time if necessary).  This equation implements the conservation of energy and can be viewed as a constraint equation
\ben\label{cse}
\dot T^2 - 1 + \frac{V^2}{E^2} = 0.
\een 
The equation of motion with this constraint can be obtained from the above by differentiating both sides one more time with respective to time  and we have
\ben
\ddot T = - \frac{V V'}{E^2},
\een 
where $V' \equiv d V/dT$.  This equation can be equivalently derived from the following proposed Lagrangian 
\ben
L (T, \dot T) = \frac{1}{2} \left( \dot T^2 + 1 - \frac{V^2 (T)}{E^2}\right),
\een
which can serve our purpose to give the Klein-Gordon equation for the tachyon wave function. For this, let us derive the corresponding Hamiltonian. The conjugate momentum of T from the above action is 
\ben 
p_T = \frac{\pa L}{\pa \dot T} = \dot T,
\een 
and as usual the Hamiltonian is 
\ben \label{hT}
H = p_T \dot T - L = \frac{1}{2} \left( \dot T^2 - 1 + \frac{V^2 (T)}{E^2}\right),
\een
which is nothing but proportional to the left side of the constraint equation (\ref{cse}), as expected.  In other words, $H = 0$ determines the tachyon background.  When expressed in terms of momentum $p_T$, we have 
\ben \label{HT}
H (T, p_T)  = \frac{1}{2} \left( p_T^2 - 1 + \frac{V^2 (T)}{E^2}\right), 
\een
whose consistency can be further verified from $\dot T = \pa H/\pa p_T=p_T$ and $\dot{p}_T = -\pa H/\pa T$.  We then pass $H = 0$ to the tachyon wave function $\psi (T)$ as 
\begin{equation}\label{e:effs}
 \left(\pa_T^2+1-\frac{1}{l^2\,\cosh^2 \beta T} \right)
\psi(T)=0,
\end{equation}
which can be derived from the following Klein-Gordon equation with $\phi (t, T) = e^{\pm i \omega t} \psi (T)$ ($\omega > 0$)
\begin{equation}\label{e:effKG}
 \left(- \pa_t^2 + \pa_T^2- \frac{1}{l^2\,\cosh^2 \beta T} \right)
\phi(t, T)=0,
\end{equation}
where as before $l \equiv E/\tau_p$.  Note that the tachyon wave function $\phi (t, T)$  is the one with its frequency $\omega =  1$ in proper units. Our approach adopted here is very muck like the usual derivation of Hawking radiation for a given black hole background  such as the Schwarzschild one. For large $T \to \infty$, the potential becomes vanishing and our interest is the positive-frequency outgoing mode $\phi (t, T) \sim e^{ - i (t - T)}$.  So we need  $\psi (T) \sim e^{i T}$ for large $T$. 

Following \cite{Anderson:2013ila},  we define 
$z=i\sinh(\be T)$ and $\psi=\sqrt{\cosh(\be T)}\hat{\psi}$.  Then the
equation (\ref{e:effs})  becomes the associated Legendre equation,
\begin{equation}\label{e:effKG1}
 \left[(1-z^2)\pa_z^2-2z\pa_z - \frac{k^2}{1-z^2}
-\left(m^2+\frac{1}{4}\right)\right] \hat{\psi}=0,
\end{equation}
where $k^2= - \left[1/(\be^2 l^2)-1/4\right] $ and $m=1/\be$. We consider 
the case of $E \ge 2\tau_p/\be$ so that $k$ is real. The normalized solution which is smooth across $T=0$ is given by the
Legendre polynomials of the first kind
\begin{equation}\label{e:effKGsolP}
 \psi^{(0)}(T)=e^{-\frac{ik \pi}{2}\sgn(T)}\frac{|\Ga(\frac{1}{2}
+k+im)|}{\sqrt{2}}\sqrt{\cosh(\be T)}P^{-k}_{-\frac{1}{2}- im}
(i\sinh(\be T)).
\end{equation}
We now focus on $T \ge 0$ since we are interested in tachyon condensation process (brane decay)  and for large $T$  we need $\psi (T) \sim e^{i T}$ as mentioned earlier. 
 This outgoing mode is given by the Legendre polynomials
of the second kind as,
\ben\label{e:effKGsola}
 \psi^{\rm out} (T) =  \frac{ i\, e^{\frac{\pi m}{2}}}{|\Ga(\frac{1}{2}
-k+im)|}\sqrt{\frac{\cosh(\be T)}{\sinh(\pi
m)}}Q^{-k}_{-\frac{1}{2}-im} (i\sinh(\be T)).\een
The normalization factor above is determined with the requirement of the asymptotical
form of solutions, i.e., $T \to \infty$, taking the following forms
\begin{equation}\label{e:effKGsolasym1}
 \psi^{\rm out} (T) \rightarrow\frac{(-)^k }
{\sqrt{2 m}} \,e^{i\frac{\pi}{4}}\, e^{-i\th} \, e^{iT} 
\end{equation}
where $\th=\arg[\Ga(1-im)\Ga(1/2-k+im)]$. The outgoing mode can be related to $\psi^{(0)} (T)$ and $\psi^{(0)*} (T)$ given in (\ref{e:effKGsolP}) as
\begin{equation}\label{e:Bog1}
 \psi^{\rm out} (T) = b^* \psi^{(0)}+ a^* \psi^{(0)*},
\end{equation}
where $*$ denotes the complex conjugate and  the Bogoliubov coefficients $a$ and $b$, satisfying $|a|^2 - |b|^2 = 1$,  are given as 
\begin{equation}\label{e:ab}
 a=-ie^{\pi m}b^*=\frac{-ie^{\frac{ik\pi}{2}}e^{\frac{\pi m}{2}}}
{\sqrt{2\sinh(\pi m)}}. 
\end{equation}
In having (\ref{e:effKGsolP}) - (\ref{e:ab}),  we have followed \cite{Anderson:2013ila} closely and used various properties of $P^\mu_\nu (z)$ and $Q^\mu_\nu (z)$ given in 
\cite{erdelyi1953higher}.  Hence, from the Bogolubov transformations (\ref{e:Bog1}) 
 and the unitarity relation $|a|^2 - |b|^2 = 1$, the particle density seen by an observer at $T\to \infty$  in the $T = 0$ vacuum 
at frequency $\omega = 1$ is
\ben
n_{\omega = 1} = |b|^2 = \frac{1}{e^{2\pi m} - 1},
\een
where we have used the relation between $a$ and $b^*$ given in (\ref{e:ab}). So the temperature of tachyon thermal radiation from this is
\ben \label{e:tachyonr}
 T_{\rm tachyon} = \frac{1}{2\pi m}  = \frac{\beta}{2\pi},
 \een 
 where we have used in the last equality $m = 1/\beta$ defined earlier.  This temperature is actually nothing but the (Gibbons-)Hawking one of dS space which will be discussed in 
 the following subsection.  
 
\subsection{(Gibbons-)Hawking and Hagedorn temperatures}

The Gibbons-Hawking temperature for dS universe can be read from the period of Euclideanized time in  (\ref{e:closeddSu}) or (\ref{e:flatdSu}) 
as
\ben \label{ght}
H^{(t)}_{\rm GH} = \frac{\beta}{2\pi}.
\een
This temperature and the Hawking temperature (\ref{e:Hawkingtemp}) in static dS space 
 can both be derived also by quantum field theory in the corresponding dS
spacetime. In the previous subsection, we have derived the thermal radiation from the tachyon condensation with a temperature (\ref{e:tachyonr}). This temperature of tachyon radiation is clear to be the same as the corresponding Hawking or Gibbons-Hawking temperature of respective dS space, 
\begin{equation}\label{e:}
 T_{H}^{(\eta)}=T_{tachyon}^{(\eta)},
\textrm{ }\textrm{ }\textrm{ } T_{GH}^{(t)}=T_{tachyon}^{(t)}.
\end{equation}
if the corresponding time is identified. This identification of two kinds of temperatures is consistent with the dynamical connection of dS space to tachyon field theory revealed previously and naturally implies that the thermal radiation in dS space is related to that in the tachyon field theory.  

As discussed in section 2, in string theory context, $\beta$ is related to string scale and in string units, the 
 temperature of tachyon radiation derived in the previous subsection is nothing but the Hagedorn temperature given in Eq.\ (\ref{e:Hagedorntemp}) if 
 the time $t$ used in the effective field theory is identified with the
zero-mode time coordinate $x^0$ in the BCFT. The Hagedorn temperature in tachyon field theory indicates a phase transition of decaying branes to closed strings. Then the natural question is how to understand its correspondence and the underlying picture in dS space, i.e., in static dS and expanding dS.  We try to address this in the following subsection

\subsection{The interpretation of Hagedorn transition in dS space}
\label{sec:fate}
The appearance of Hagedorn temperature in the thermal radiation of tachyon field theory indicates a transition to closed strings. Closed string emissions from rolling tachyon have been investigated in previous works. It is found that rolling tachyon can decay into
massless \cite{Chen:2002fp,Rey:2003xs} and massive
\cite{Okuda:2002yd,Lambert:2003zr} closed strings. The calculations
in BCFT in \cite{Lambert:2003zr} suggest that homogeneously decaying
unstable D$p$-brane with $(p\leq2)$ wrapped on a sphere can
completely decay into closed strings, mainly massive ones, at the
end of tachyon condensation. 

What is then the underlying picture of this transition when looked from the bulk dS space?  The inevitable existence of a (cosmological) event horizon in dS space is due to the fact that no single observer can access the entire dS spacetime. The consequence of this is that dS space is naturally associated with a temperature, as shown in \cite{Spradlin:2001pw}, 
which is the same for any observer moving along a time-like geodesic and is determined by the dS radius $r_H$.  In other words, any geodesic observer in dS space will feel that she/he is in a thermal bath of particles at a temperature 
\ben \label{dstemp} 
T_{\rm dS} = \frac{1}{2\pi r_H},
\een
where the dS radius $r_H = 1/\beta$ with $\beta$ related to the cosmological constant as given below (\ref{e:stadS}).  So for concreteness and without loss of generality, we will focus on a specific observer stationary at the south pole and consider the static dS space as given by the metric (\ref{e:stadS}). For this observer, the region fully accessible cannot go beyond the horizon. 

If the underlying theory is indeed string theory, then from section 2, we know that $r_H$ is on the order of string length $l_s$. Then the physical radius of the $(d - 1)$-sphere in the metric   (\ref{e:stadS}) is just $r \in [0, r_H] \lesssim l_s$, which is also on the order of string length. The physical distance $L$ from $r = 0$ to the horizon $r = r_H$ can be calculated to be $L = \pi r_H/2 \sim l_s$, again also on the order of string length.  In other words, when string theory is considered, the region fully accessible to the observer has a physics size of string length.  For such small physical size, as raised in section 2, whether the dS geometry can still be taken classically is certainly questionable but nevertheless the above obtained temperature (\ref{dstemp}) still agrees with more sophisticated stringy computations\cite{Strominger:2002pc,Maloney:2003ck}. So we may assume that the dS geometry can still be good or at least on the verge of breakdown. With this, we can now address the question posed at the end of the previous subsection. First, the thermal temperature given in (\ref{dstemp}) agrees now with the Hagedorn temperature (\ref{e:Hagedorntemp}).  If the dS geometry originates indeed from string theory, we have then only closed strings in the bulk dS spacetime. But to the observer at the south pole, the accessible physical size to him/her is on the order of string length. Therefore, the observer can most likely see part of the closed strings at such a high temperature and the rest part of them should lie beyond the horizon and cannot be seen by the observer.  In other words, the closed strings appear to the observer as if they were open strings and this may explain the connection of dS dynamics to that of worldvolume tachyon field theory at low energy. 

We are now ready to provide a dS picture of Hagedorn transition. In tachyon field theory, the appearance of Hagedorn temperature signals a phase transition of open string modes to closed strings. In the bulk dS space, there exist only closed strings but they appear to the observer as if they were open strings since the accessible region to any given observer is limited to one enclosed by its horizon, whose size is on the order of string scale, due to the peculiar property of dS space. At such a high temperature, the seemingly open strings will escape the region accessible to the observer. Once this happens, these open strings disappear with respect to the observer and they restore what they are as closed strings. This process may be thought, with respect to the observer, as a transition process of open strings to closed strings even though the observer doesn't know this transition.  This is consistent with the world-volume picture of Hagedorn transition for which the world-volume dynamics itself doesn't imply this transition in the $g_s = 0$ limit and it is the appearance of Hagedorn temperature to signal this to happen. An alternative picture of this is to assume that the classical dS geometry is on the verge of breakdown at the Hagedorn temperature. Before the breakdown,  the observer can only see open strings in the region accessible to him/her. When the breakdown happens, the dS geometry ceases to exist and the peculiar property of dS geometry is lost.  As such, the immortal observer can access much large region of space and see the previously seemingly open strings to restore to what they are as closed strings.

\section{Conclusion}
\label{sec:conculsions}

In this paper, we have provided evidence supporting that the (thermo)dynamics in dS space can be described 
by the tachyon field theory of unstable D-branes. Concretely,  we have shown that the radial geodesic motion of a massive probe particle
in static dS space can be described by an action, which  turns out to be exactly the same as that of the tachyon field theory derived from open string field theory. Further we have also 
shown that certain low energy linear scalar dynamics in the global or flat dS space can be identified with the tachyon fluctuations on a homogeneous tachyon background, representing either full S-brane or half S-brane, respectively.  In addition, we have also shown that the thermal temperature of tachyon radiation agrees with that felt by any time-like observer in dS space.  In particular, in string theory context, this temperature is actually the Hagedorn one, signaling a transition to closed strings.  An understanding of this transition in dS space is also proposed and it is believed to be due to, in the present context,  that the size of the region fully accessible to any timelike observer in dS space is on the order of string scale or the dS space ceases to exist, when the thermal temperature reaches the Hagedorn temperature.   

The present study also provides an understanding of the tachyon condensation geometrically in terms of dS dynamics.  For example, the tachyon rolling from the top of tachyon potential
to the bottom can be represented by the geodesic motion of a massive particle in a static dS space from, say, the south pole to the event horizon.  The open string vacuum corresponds to just the near south pole geometry, which is a flat Minkowski, while the closed string vacuum to the near horizon geometry, which is now a Rindler spacetime times  a sphere. In the present context, the transition from the open string vacuum to the closed string one can be geometrically represented by the static dS geometry, which interpolates the flat Minkowski to the Rindler space times a sphere.  

For $dS_d$  space in global coordinates with $d > 1$,  except for the $d = 3$ case,  we can identify, only for a very limited modes, a probe scalar in this dS background with the corresponding tachyon fluctuations in certain homogeneous full-S brane background on unstable Dp-branes wrapped on a p-sphere. We are presently lack of an understanding       
 of this limitation except for the speculation mentioned at the end of subsection 3.1. We try to address this elsewhere in the near future. 

\section*{Acknowledgements\markboth{Acknowledgements}{Acknowledgements}} 
HL is grateful for the hospitality of Kavli Institute for
Theoretical Physics China, CAS, and Interdisciplinary Center for Theoretical Study, USTC, where part of the work has been done.  JXL acknowledges support by a key grant from the NSF of China with Grant No : 11235010.
\bibliographystyle{JHEP}
\bibliography{b}

\end{document}